\documentclass[aps,prl,showpacs,twocolumn]{revtex4} 
\usepackage{times}
\usepackage{epsfig}

\begin{document}

\title{Non-Markovian Particle Dynamics in Continuously Controlled
  Quantum Gases}

\author{D.~Ivanov}

\affiliation{Fachbereich Physik, Universit\"at Rostock,
  Universit\"atsplatz 3, D-18051 Rostock, Germany}

\author{S.~Wallentowitz}

\affiliation{Fachbereich Physik, Universit\"at Rostock,
  Universit\"atsplatz 3, D-18051 Rostock, Germany}

\date{\today}

\begin{abstract}
  For a quantum gas, being subject to continuous feedback of a
  macroscopic observable, the single-particle dynamics is studied.
  Albeit feedback-induced particle correlations, it is shown that
  analytic solutions are obtained by formally extending the
  single-particle Hilbert space by an auxiliary degree of freedom.
  The particle's motion is then fed by colored noise, which
  effectively maps quantum-statistical correlations onto the single
  particle. Thus, the single particle in the continuously controlled
  gas follows a non-Markovian trajectory in phase-space.
\end{abstract}


% 05.40.-a  Fluct phenomena, ..., Brownian motion.
% 05.30.-d  Quantum statistical mechanics
% 05.30.Jp  Boson systems (for Bose-Einstein condensation, see
%           03.75.Fi)

\maketitle

As noted by Caves and Milburn~\cite{caves-milburn}, the continuous
observation of an object inevitably results in an increasing
uncertainty of its momentum, which may cause it to eventually escape
from the region of observation.  An experiment thus necessarily
requires a mechanism to keep the object at fixed location in the
laboratory. Such constraining forces can be covered under the general
notion of feedback control.

Given limited observational capabilities, in most cases the
macroscopic body is observed as a whole, its internal structure
remaining unresolved. Clearly the same holds also for a continuous
feedback process, where measurements and conditioned unitary actions
follow in continuous sequence. In consequence the issue may be raised,
how the microscopic constituents of the object are affected by the
feedback.

It is well known that macroscopic systems with large numbers of
correlated constituents allow for exact solutions only for few model
systems~\cite{korepin}. The correlations, being responsible for the
large degree of complexity, are usually generated by interactions
between the constituents. Since internal and external degrees of
freedom are usually considered as being decoupled, one might think
that macroscopic observations have no effect on the internal dynamics
of a system~\footnote{The inverse relation, e.g. having coherence in
  the center of mass of fullerenes despite internal degrees of freedom
  being excited~\cite{zeilinger} may suggest this viewpoint too.}.
However, strong and high-order correlations emerge also due to
measurement or feedback control of a macroscopic
observable~\cite{wallentowitz}.

In this Letter we show, that substantial effects on the particles of a
continuously feedback-controlled quantum gas emerge. We prove, that,
despite the above mentioned problems of correlated systems, continuous
feedback can be exactly solved for the single-particle dynamics. This
dynamics contains effects due to particle correlations, that are
formally introduced as colored noise feeding the single-particle
motion.

Continuous measurement~\cite{measurement,caves-milburn} and
feedback~\cite{wiseman-milburn,feedback} have been studied and
experimentally realized~\cite{rempe} in the past for various systems,
such as single atoms or single harmonic systems.  Different from these
treatments, we address here the case of a bosonic many-particle system
and thus deal with the additional complexity due to particle
correlations.

Consider an ideal gas where the constituents of mass $m$ are
indistinguishable, bosonic particles. In order to keep at all times
the center of the cloud of particles at a predefined target position,
a feedback loop is continuously applied to compensate for the motion
of the center of mass of the gas. The feedback will act as a damping
force and unavoidably also randomizes the motion of the center of
mass.
%, see Fig.~\ref{fig:scheme}.

%% \begin{figure}
%%   \centering
%%   \includegraphics[width=0.4\textwidth]{figure1.eps}    
%%   \caption{The center of mass $X$ is continuously observed and kicked
%%     back, resulting in Brownian motion of $X(t)$. The single particle
%%     will also perform a stochastic motion $x(t)$, whose type is to be
%%     determined.}
%%   \label{fig:scheme}
%% \end{figure}

The quantum dynamics due to the continuous feedback process has been
shown~\cite{caves-milburn} to be governed by the master equation of quantum
Brownian motion~\cite{brownian1,brownian2},
\begin{eqnarray}
  \label{eq:N-master}
  \dot{\hat{\varrho}} 
  & = & - \frac{i}{\hbar} [ \hat{H}, \hat{\varrho}
  ]  + i \frac{\zeta}{2\hbar} [ \hat{P}, \{ \hat{X}, \hat{\varrho}
  \} ] \nonumber \\
  & & - \frac{1}{8\sigma^2} [ \hat{X}, [\hat{X}, \hat{\varrho}]]
  - \frac{\zeta^2\sigma^2}{2\hbar^2} [ \hat{P}, [ \hat{P},
  \hat{\varrho}]]. 
\end{eqnarray}
Here $\hat{H}$ is the Hamiltonian of the free evolution of the many-particle
system, which we specify as linear dynamics of an ideal gas, and the
additional terms are due to the feedback.

Continuous feedback is obtained as the limit of a discrete series of
measurements and kicks of the center of mass $\hat{X}$: In each
discrete step $\hat{X}$ is measured with resolution $\sigma_0$ and
then a kick is applied leading to a shift $-\zeta_0 X$, where $X$ is
the measured value. After that the free evolution takes place and the
sequence is repeated.  The continuous limit, where the average rate of
feedbacks $\gamma \!\to\!  \infty$, then requires that $\sigma_0
\!\to\!  \infty$ and $\zeta_0 \!\to\! 0$, such that $\sigma \!=\!
\sigma_0/ \sqrt{\gamma}$ and $\zeta \!=\!  \zeta_0 \gamma$ remain
constant. Thus the parameters $\sigma$ and $\zeta$ in
Eq.~(\ref{eq:N-master}) describe the strengths of measurement and
subsequent kick.

It should be emphasized that for the gas considered here,
Eq.~(\ref{eq:N-master}) describes the dynamics of the {\em
  many-particle} density operator of the system $\hat{\varrho}$.
Needless to say, that for the complete many-particle problem neither
exact analytical nor numerical solutions exist. Nevertheless, it is
straightforward to obtain solutions for collective variables, such as
the center of mass $\hat{X}(t)$ or the total momentum $\hat{P}(t)$. In
fact, these properties will lack any features due to the many-particle
aspect, but will be identical to that of a single-particle
system~\cite{caves-milburn} of correspondingly larger mass.

The center-of-mass rms deviation of a harmonically trapped gas with trap
frequency $\omega_0$, for instance, will converge to the stationary value
\begin{equation}
  \label{eq:DX}
  \lim_{t\to\infty} \Delta X(t)
  = \Delta X_0 \sqrt{ \frac{\eta + \eta^{-1}}{2} } .
\end{equation}
Here $\Delta X_0 \!=\! \sqrt{\hbar/(2 M \omega_0)}$ is the
ground-state width of the center of mass (total mass $M$) and $\eta
\!=\!  (\Delta X_0)^2 / (\zeta\sigma^2)$ denotes the ratio of spatial
localization due to the potential over that due to the feedback.  The
stationary value~(\ref{eq:DX}) is typically smaller than the initial
value and thus indicates the gain in localization due to the feedback.

Different from a consideration of the controlled variables themselves,
in this Letter we address a more subtle question. It is the question,
how the {\em single particle} in the gas is affected by the continuous
feedback.
%, cf.~Fig.~\ref{fig:scheme}. 
Various important properties of the gas, that may be derived from the
single-particle behavior, for example its density profile, justify
this approach.

The single-particles dynamics can in principle be deduced from
Eq.~(\ref{eq:N-master}), accompanied with the conceptual difficulties
due to many-particle correlations. We proceed instead by mapping these
correlations onto single-particle fluctuations, which will be shown to
allow for analytical solutions.

Let us first consider the feedback in more detail in a second-quantized
picture, using the bosonic matter field $\hat{\phi}(x)$ with commutator
relation $[\hat{\phi}(x), \hat{\phi}^\dagger(x')] \!=\! \delta(x \!-\! x')$.
The total momentum of particles in the gas, as used in
Eq.~(\ref{eq:N-master}), then reads
\begin{equation}
  \label{eq:P-def}
  \hat{P} = -i \hbar \int \! dx \, \hat{\phi}^\dagger(x) \,
  \frac{\partial}{\partial x} \, \hat{\phi}(x) . 
\end{equation}
It should be noted, that the operations of measurement and kick are typically
implemented by use of external (e.g. optical) probe and control fields, that
equally interact with all particles. Thus, the accessible properties of the
gas are necessarily of extensive (i.e. additive) type.

The center of mass of the gas, however, is an intensive quantity: It
is the ratio of summed particle positions over the particle number. In
conclusion this variable cannot be accessed in the above mentioned
way. Instead, a priori information on the particle number is required
for rendering $\hat{X}$ quasi into an extensive quantity.  Typically,
the best information one can get on the particle number from some
previous measurements may be its expectation value $\langle \hat{N}
\rangle$.  Using therefore this average particle number, the truly
accessible (quasi) center-of-mass reads~\footnote{Note, that if a
  truly extensive quantity would be measured, this feature would
  appear instead in the kick operation and thus cannot be avoided in
  principle~\cite{wallentowitz}.}
\begin{equation}
  \label{eq:Q-def}
  \hat{X} = \frac{1}{\langle \hat{N} \rangle} 
  \int \! dx \, \hat{\phi}^\dagger(x) 
  \, x \, \hat{\phi}(x) .
\end{equation}

Our goal is to gain information on the dynamics of a single particle
in the gas, described by the time evolution of the reduced
single-particle density matrix,
\begin{equation}
  \label{eq:sp-dens-mat}
  \rho(x,x') = \big\langle \hat{\phi}^\dagger(x') \, \hat{\phi}(x) 
  \big\rangle .
\end{equation}
Deriving the equation of motion of this density matrix from
Eq.~(\ref{eq:N-master}), an infinite hyrarchy of coupled equations for
particle correlations of increasing order would be obtained. This
would not allow for a closed equation of motion for the
single-particle density matrix~(\ref{eq:sp-dens-mat}) alone. Of
course, this is the central problem in many-particle physics, where
only few model systems provide analytical solutions~\cite{korepin}. In
general one is forced to rely on approximations, such as the
truncation of higher-order correlations.

In our case, the many-particle correlations do not decay with
increasing order, they represent the strong correlating effects of the
measurement of the collective variable $\hat{X}$.  Any outcome of the
measurement can be realized by a vast number of microstates compatible
with the observed value $X$. After the measurement all these
microstates become parts of the highly correlated many-particle
superposition state. In addition, the indistinguishability of bosonic
particles leads to further quantum-statistical effects in this
conditioned quantum state.

There have been solutions to continuous-feedback problems for many-particle
systems, in effect however, considering distinguishable
particles~\cite{thomsen}. In this letter we deal with bosons, requiring thus a
conceptually different approach. One should note, that due to the
indistinguishability, Eq.~(\ref{eq:sp-dens-mat}) is the expectable behavior of
a representative particle in the gas and clearly not that of a specific
(distinguished) one.

Given the above mentioned problems, we proceed by extending the
Hilbert space beyond that of $\rho(x,x')$, such as to obtain closed
equations of motion.  A replacement for Eq.~(\ref{eq:sp-dens-mat}) is
searched for that, however, contains the complete information of
$\rho(x,x')$ and even more. A suitable candidate for such an approach
is the following function
\begin{eqnarray}
  \label{eq:wigner}
  \lefteqn{ W_N(x,p; X,P) = 
    (2\pi\hbar)^{-4}
    \! \int \!  dx' \! \int \! dX' \! \int \! dP' \! \int \! dS   \,
  }  & & \nonumber \\
  & & \times  \, \exp \!\left\{-i \left[ x'p \!+\! PX' \!+\! XP' \!+\!
    (N \!-\! 1) S  \right] / \hbar \right\}
  \nonumber \\
  & & \times \left\langle \hat{\phi}^\dagger \!\Big( x \!-\!
    \frac{x'}{2} \Big) \,
    e^{ i ( \hat{P} X' + \hat{X}  P' + \hat{N} S ) / \hbar }  
    \hat{\phi} \!\Big( x \!+\! \frac{x'}{2} \Big) \right\rangle . 
\end{eqnarray}
It can be interpreted as the Wigner function of a joint two-body
system, conditioned on $N$ particles being in the gas. The first body
of the system is given by a single particle with phase-space variables
$x$ and $p$.  The second one is the center of mass of the ``other''
$N\!-\!1$ particles with phase-space variables $X$ and $P$.

Tracing Eq.~(\ref{eq:wigner}) over the macroscopic variables $N$, $X$,
and $P$, the single-particle Wigner function, being equivalent to
$\rho(x,x')$, is obtained.  However, integrating over $x$, $p$ one
does not obtain the true center-of-mass Wigner function. For a large
average particle number $\langle \hat{N} \rangle \!\gg\! 1$, though,
the annihilation of only a single particle by the field operators in
Eq.~(\ref{eq:wigner}) can be neglected. Then Eq.~(\ref{eq:wigner})
converges to the true two-body Wigner function of single particle and
center-of-mass system.

We can show that for Eq.~(\ref{eq:wigner}) a closed equation of motion
can be derived from Eq.~(\ref{eq:N-master}). In consequence, also the
single-particle state $\rho(x,x')$ can be obtained by tracing over the
macroscopic variables $N, X, P$. For particles of mass $m$, being
bound in a harmonic potential of frequency $\omega_0$, for example,
$W_N(x,p;X,P)$ obeys the following Fokker--Planck equation
\begin{eqnarray}
  \label{eq:fokker-planck}
  \lefteqn{ \Bigg\{ \partial_t   
  + \frac{p \, \partial_x}{m} 
  -  m \omega_0^2 x  \, \partial_p
  + \frac{P \, \partial_X}{M} 
  - M \omega_0^2 X \, \partial_P } & & \nonumber \\
  & & 
  - \frac{\hbar^2}{8 \sigma^2} \bigg( \Theta_N \partial_P \!+\! 
    \frac{\partial_p}{\langle \hat{N} \rangle} \bigg)^2
 -  \frac{\zeta^2 \sigma^2}{2} ( \Theta_N \partial_X 
  \!+\! \partial_x )^2 \nonumber \\
  & & 
  - \zeta ( \Theta_N \partial_X \!+\! \partial_x ) 
  \bigg( X \!+\! \frac{x}{\langle \hat{N} \rangle} 
  \bigg) \Bigg\} \, W_N(x,p; X,P) = 0 . \qquad
\end{eqnarray}
Here $\Theta_N \!=\! (N \!-\! 1) / \langle \hat{N} \rangle$, the total
mass is $M \!=\! m \langle \hat{N} \rangle$, and $N$ plays the role of
a parameter.  This Fokker--Planck equation is of linear type with a
positive semi-definite diffusion matrix. Thus a bound analytic
solution for its Green function of Gaussian type can be
found~\cite{risken}. In consequence, the analytic solution for
$\rho(x,x')$ can be obtained in a rather straightforward way from the
(analytic) solution of Eq.~(\ref{eq:fokker-planck}).

For getting more insight into the dynamics described by
Eq.~(\ref{eq:fokker-planck}), we now turn to the equivalent
$N$-parameterized stochastic differential equations, that read
\begin{eqnarray}
  \label{eq:x}
  dx_N & = & \bigg[ \frac{p_N}{m} \!-\! \zeta \bigg( \frac{x_N}{\langle
    \hat{N} \rangle} \!+\! X_N \bigg) \bigg] dt 
  +  \zeta \sigma \, d\xi_1 , \\
  \label{eq:p}
  dp_N & = & -m \omega_0^2 x_N \, dt + 
  \frac{\hbar}{2 \sigma \langle \hat{N} \rangle} \, d\xi_2 , \\
  \label{eq:X}
  dX_N & = & \bigg[ \frac{P_N}{M} \!-\! \zeta \Theta_N 
    \bigg( \frac{x_N}{\langle \hat{N} \rangle} \!+\! X_N \bigg) \bigg] dt
  \!+\!  \Theta_N \zeta \sigma \, d\xi_1 , \quad \\
  \label{eq:P}
  dP_N & = & - M \omega_0^2 X_N \, dt +
    \frac{\hbar\Theta_N}{  2 \sigma} \, d\xi_2  , 
\end{eqnarray}
with $\xi_1$, $\xi_2$ being statistically independent Wiener
processes.

Besides the free evolution of the single particle and of the (quasi)
center of mass, these equations also contain a feedback-induced
coupling between the microscopic single-particle and the macroscopic
center-of-mass degree of freedom. Thus our method of solution, that
incorporates all many-particle correlation effects in the
single-particle dynamics, was obtained by allowing for a coupling with
an auxiliary macroscopic degree of freedom.

It is important to note, that since the drift matrix of
Eqs~(\ref{eq:x})--(\ref{eq:P}) is not of normal form, it is impossible
to diagonalize this set of dynamical equations.  Furthermore, both
systems are fed by the same noise sources, which is apparent since
both macroscopic and microscopic systems are subject to the same
feedback process.

When tracking the motion of both systems, i.e. microscopic {\it and}
macroscopic, Eqs~(\ref{eq:x})-(\ref{eq:P}) generate Markovian
trajectories in the four-dimensional phase space. The macroscopic
system, however, being the center of mass of the $N \!-\! 1$ ``other''
particles, is not a physically accessible observable. Thus it appears
natural to eliminate this auxiliary degree of freedom, to obtain
stochastic differential equations for the microscopic trajectory
alone. Let us perform this elimination for the case of large particle
numbers:

For $\langle \hat{N} \rangle \!\to\! \infty$, due to its much larger
inertia, the (quasi) center-of-mass system will no longer be affected
by the motion of the single particle. Thus in Eq.~(\ref{eq:X}) the
term proportional to $x_N/\langle \hat{N} \rangle$ can be discarded,
which decouples the macroscopic system. Consequently, the formal
solution $X_N(t)$ is then obtained as Ornstein--Uhlenbeck process.
Inserting it into Eq.~(\ref{eq:x}), one is finally left with equations
of motion for $x_N$ and $p_N$ alone.

The modifications of Eq.~(\ref{eq:x}) due to the elimination procedure
are formally done by two replacements: $X_N$ in Eq.~(\ref{eq:x})
becomes the deterministic part of the solution $X_N(t)$, which now
acts as an external drive. And the Wiener increment $d\xi_1(t)$ is
replace by $d\xi_N(t)$, which is now a sum of a Wiener and an
Ornstein--Uhlenbeck process:
\begin{eqnarray}
  \label{eq:new-noise}
  \lefteqn{ d\xi_N(t) = d\xi_1(t) - 2 \Gamma_N dt \int_0^t 
   \bigg\{  d\xi_1(t') \cos[\Omega_N (t\!-\! t')] } & & 
  \nonumber \\
  & & + \,
  \frac{\eta \, d\xi_2(t') \!-\! \alpha_N \, d\xi_1(t')}{\sqrt{1 \!-\!
      \alpha_N^2}} \sin[\Omega_N (t \!-\! t')] \bigg\} e^{-\Gamma_N(t-t')} . 
  \qquad
\end{eqnarray}
Here $\Omega_N^2 \!=\! \omega_0^2 \!-\!  \Gamma_N^2$ and $\alpha_N \!=\!
\Gamma_N / \omega_0$, with $\Gamma_N \!=\!  \zeta \Theta_N / 2$ being the
feedback-induced damping of the coherent oscillation in the trap potential.

\begin{figure}
  \centering
  \includegraphics[width=0.45\textwidth]{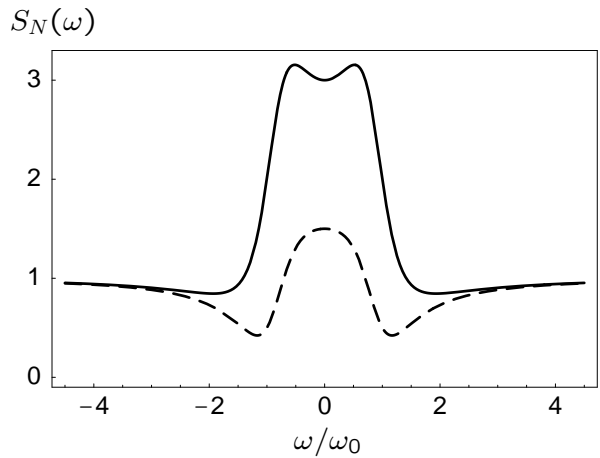}   
  \caption{Noise spectrum for $\alpha_N \!<\! 1 / \sqrt{2}$ ($\alpha_N
    \!=\! 1/2$) for the case satisfying Eq.~(\ref{eq:eta-condition})
    with $\eta \!=\! \sqrt{2}$ (solid), and for $\eta \!=\!  1 /
    \sqrt{2}$ (dashed).}
  \label{fig:noise1}
\end{figure}

\begin{figure}
  \centering
  \includegraphics[width=0.45\textwidth]{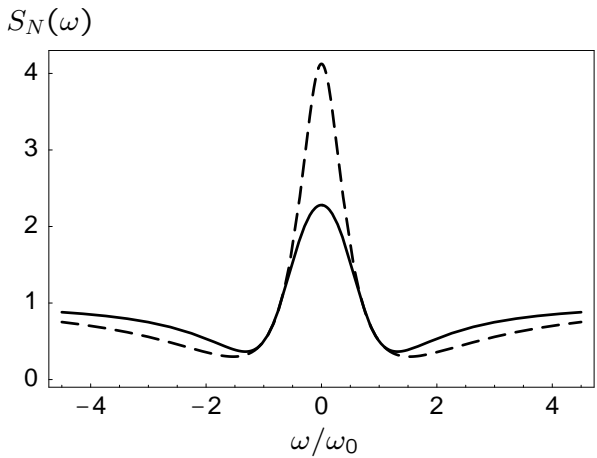}    
  \caption{Noise spectrum for $\alpha_N \!>\! 1 / \sqrt{2}$ ($\eta
    \!=\! 1 / \sqrt{2}$), for the oscillatory case $\alpha_N \!=\!
    0.8$ (solid) and the over-damped case $\alpha_N \!=\! 1.25$
    (dashed).}
  \label{fig:noise2}
\end{figure}

To characterize this effective noise source we consider the Fourier
transform of the stationary correlation function
\begin{equation}
  \label{eq:spectrum}
  S_N(\omega) = \lim_{t\to\infty} \int \! d\tau \,
  e^{i\omega\tau} \, \overline{\dot{\xi}_N(t \!+\! \tau)
  \dot{\xi}_N(t)} ,
\end{equation}
which is a symmetric spectrum depending on $N$ only via the parameter
$\alpha_N$. From Figs~\ref{fig:noise1} and \ref{fig:noise2} it is
observed that only in the high-frequency limit the white-noise
background $S_N(\omega) \!=\!  1$ is reached. At low frequencies
several extrema appear depending on the values of $\alpha_N$ and
$\eta$ (or instead $\zeta$).

For weak feedback-induced damping, $\alpha_N \!<\! 1 / \sqrt{2}$, the
peaks at the damped center-of-mass oscillation frequencies
$\pm\Omega_N$ can be resolved under the condition
\begin{equation}
  \label{eq:eta-condition}
  \eta > 1 / \sqrt{ 2 - 4 \alpha_N^2} ,
\end{equation}
see Fig.~\ref{fig:noise1}. Such large values of $\eta$ represent
strong feedback-induced localization of the center of mass. For
smaller values of $\eta$ the two peaks overlap and only a single
maximum at $\omega \!=\! 0$ remains (see dashed curve in
Fig.~\ref{fig:noise1}). A remarkable effect, however, appears as
minima are formed at frequencies larger than $\omega_0$.  These minima
represent noise reduction below the white-noise background and are
generated by self-correlation of white-noise with its modulated
version, as seen from the contributions to Eq.~(\ref{eq:new-noise}).

For $\alpha_N \!\geq\! 1/\sqrt{2}$, i.e. for large damping, the maxima
at $\pm\Omega_N$ can never be resolved, leaving the single maximum at
$\omega \!=\! 0$. Also in this case symmetric minima at noise levels
below unity appear at frequencies slightly higher than the trap
frequency $\omega_0$, cf.~Fig.~\ref{fig:noise2}.

This noise reduction may be understood as follows: Both the single
particle and the (quasi) center of mass are subject to the same
feedback-generated noise. With some time delay the center of mass
transfers its noise to the single particle via the coupling,
cf.~Eqs~(\ref{eq:x}) and (\ref{eq:X}). Thus the effective noise seen
by the single particle at the frequencies of the minima is partially
compensated due to destructive phase shifts of the two paths of the
noise input.

Since the noise feeding the single-particle coordinate is now colored,
as seen from the previous discussion, the resulting stochastic
trajectories in single-particle phase space will be {\it
  non-Markovian}. This feature is due to the feedback-generated strong
many-particle correlations, that are now cast into an effective noise
source for the single particle. Thus the coloredness of the
noise~(\ref{eq:new-noise}) is a manifestation of generation of
correlations between the particles.

One may interpret this also as a feedback-mediated effective
interaction of the single particle with the cloud of surrounding
particles, that bears a non-vanishing correlation time.  The latter
may be quantified by the inverse spectral width of
Eq.~(\ref{eq:spectrum}). Thus the memory effects in the particle cloud
can be made responsible for the non-Markovian trajectory of the single
particle.

Summarizing, we have shown that a specific many-particle problem, that
is dominated by strong particle correlations can be analytically
solved for the reduced single-particle dynamics. The solution was
obtained by first extending the Hilbert space of a single particle by
an auxiliary degree of freedom to obtain analytically solvable closed
equations of motion. Eliminating the auxiliary degree of freedom, it
could be shown that the single particle follows a non-Markovian
trajectory in phase space. Thus the many-particle correlations have
been mapped onto a coloredness of the noise feeding the particle's
motion.

The specific problem, for which this solution was obtain is quite
general, in that it represents typical control of a system on the
macroscopic level. In conclusion even such a macroscopic intervention
creates correlations and a resulting non-Markovian behavior of single
particles inside the controlled system.

One may ask for a generalization of our approach: Is it generally
possible to map many-particle correlations onto specific features of
noise sources feeding the motion of single particles? Clearly in the
presented case, the correlations were not generated by true particle
interactions but by continuous feedback. Thus an immediate application
to problems of interacting particles seems not obvious.  Considering
however, that we can interpret the solved system equally well as
quantum Brownian motion of scattering, and thus interacting,
particles, a potential route to a more general methodology may come
into sight.

This research was supported by Deutsche Forschungsgemeinschaft.

\end{document}